\documentclass[prl,twocolumn,showpacs,superscriptaddress]{revtex4}

\usepackage{amsmath,amssymb}
\usepackage{verbatim}
\usepackage{graphicx}
\usepackage{hyperref}
\usepackage{color}

\DeclareFontFamily{OT1}{rsfs}{}
\DeclareFontShape{OT1}{rsfs}{m}{n}{ <-7> rsfs5 <7-10> rsfs7 <10->rsfs10}{} 
\DeclareMathAlphabet{\mycal}{OT1}{rsfs}{m}{n}

\let\Lw\L
\renewcommand{\L}{{\mathcal{L}}}

\newcommand{\be}[1]{ \begin{equation}\label{#1} }
\newcommand{\ee}{\end{equation}}
\newcommand{\bea}[1]{\begin{eqnarray}\label{#1} }
\newcommand{\eea}{\end{eqnarray}}

\newcommand{\eq}[2]{\begin{equation} #1 \label{#2} \end{equation}}

\newcommand{\de}{\delta}

\begin{document}

\title{Higher spin theory in 3-dimensional flat space}

\author{Hamid Afshar}
\email{afshar@hep.itp.tuwien.ac.at}
\affiliation{Institute for Theoretical Physics, Vienna University of Technology, Wiedner Hauptstrasse 8--10/136, A-1040 Vienna, Austria}

\author{Arjun Bagchi}
\email{arjun.bagchi@ed.ac.uk}
\affiliation{School of Mathematics, University of Edinburgh, Edinburgh EH9 3JZ, United Kingdom}
\affiliation{Indian Institute of Science Education and Research (IISER), Pune, Maharashtra 411008, India}

\author{Reza Fareghbal}
\email{fareghbal@theory.ipm.ac.ir}
\affiliation{Institute for Theoretical Physics, Vienna University of Technology, Wiedner Hauptstrasse 8--10/136, A-1040 Vienna, Austria}
\affiliation{School of Particles and Accelerators, Institute for Research in Fundamental Sciences (IPM), P.O. Box 19395-5531, Tehran, Iran}

\author{Daniel Grumiller}
\email{grumil@hep.itp.tuwien.ac.at}
\affiliation{Institute for Theoretical Physics, Vienna University of Technology, Wiedner Hauptstrasse 8--10/136, A-1040 Vienna, Austria}

\author{Jan Rosseel}
\email{rosseelj@hep.itp.tuwien.ac.at}
\affiliation{Institute for Theoretical Physics, Vienna University of Technology, Wiedner Hauptstrasse 8--10/136, A-1040 Vienna, Austria}

\date{\today}

\preprint{TUW--13--xx}

\begin{abstract} 
We present the first example of a non-trivial higher spin theory in 3-dimensional flat space.
We propose flat-space boundary conditions and prove their consistency for this theory.
We find that the asymptotic symmetry algebra is a (centrally extended) higher spin generalization of the Bondi--Metzner--Sachs algebra, which we describe in detail.
We also address higher spin analogues of flat space cosmology solutions and possible generalizations.
\end{abstract}


\maketitle

Flat space is a good approximation for most purposes in physics, including particle physics.
In flat space the possible spins of interacting fundamental particles are highly constrained by consistency requirements.
In fact, various no-go results forbid non-trivial (massless) higher spin theories in 4- or higher-dimensional flat space. 
These no-go results include the Coleman--Mandula theorem \cite{Coleman:1967ad}, the Aragone--Deser no-go result \cite{Aragone:1979hx}, the Weinberg--Witten theorem \cite{Weinberg:1980kq} and others, see \cite{Bekaert:2010hw} for a nice summary.
Consistently, all the observed fundamental particles so far have either spin 1 (all the gauge bosons), spin $\tfrac 12$ (all fermionic matter fields) or spin 0 (the Higgs boson).

Of course, every no-go theorem is only as good as its premises. For instance, the Coleman--Mandula theorem is circumvented by supersymmetry, which then leads to a generalization in form of the Haag--\Lw opusza\'{n}ski--Sohnius theorem \cite{Haag:1974qh} and to many interesting theoretical and phenomenological applications. Vasiliev pioneered another way to circumvent all the no-go theorems mentioned above by considering higher spin theories in (constantly) curved spacetimes \cite{Vasiliev:1990en}.
This seminal work recently has led to numerous research activities due to the holographic conjecture in four bulk dimensions about a decade ago \cite{Klebanov:2002ja,Sezgin:2002rt} and the more recent one by Gaberdiel and Gopakumar in three bulk dimensions \cite{Gaberdiel:2010pz}.

These conjectures, if true, establish novel kinds of correspondences between non-trivial higher spin theories and weakly coupled (conformal) field theories.
They are thus complementary to ``usual'' holography, which relates interacting gravitational theories with strongly coupled field theories \cite{Aharony:1999ti}, and allow to test holographic ideas in regimes that are otherwise inaccessible, see e.g.~\cite{Giombi:2009wh,Maldacena:2011jn,Vasiliev:2012vf,Gaberdiel:2012uj,Ammon:2012wc} for selected recent developments and reviews. It is fair to say that nearly all these developments are related to higher spin theories in spacetimes of (non-vanishing) constant curvature (see \cite{Gary:2012ms,Afshar:2012nk} for some exceptions).

Interestingly, in three spacetime dimensions none of the no-go results mentioned above apply, so that it could be possible to construct non-trivial (massless) higher spin theories in asymptotically flat space. Nevertheless, so far no example exists that realizes a non-trivial higher spin theory in 3-dimensional flat space. 
To be precise, by ``non-trivial higher spin theory in three dimensions'' we mean an analogue of Einstein gravity ($=$~spin-2 gravity), i.e., the theory is allowed to be topological in the bulk, but must have non-trivial boundary excitations, which fall into (not necessarily unitary) representations of some higher spin algebra.
The absence of physical bulk degrees of freedom as well as the general considerations on contractions of higher spin algebras in the seminal work by Fradkin and Vasiliev (see section 6 of \cite{Fradkin:1987}) make it plausible that the desired construction should work.

Indeed, in this paper we provide the first example of a higher spin theory in 3-dimensional flat space and develop its holographic description in terms of a dual field theory whose symmetry algebra turns out to be a higher spin generalization of the Bondi--Metzner--Sachs (BMS) algebra \cite{Bondi:1962,Sachs:1962,Ashtekar:1996cd,Barnich:2006av}. We also determine its possible (and its actual) central extensions. Finally, we address flat space cosmology solutions and possible generalizations.

For simplicity, we consider spin-3 gravity. In Anti-de~Sitter space (AdS) this theory is most efficiently formulated as $sl(3)_k\oplus sl(3)_{-k}$ Chern--Simons theory with bulk action ($k$ is the Chern--Simons level, the only coupling constant of the theory)
\eq{
I=\frac{k}{4\pi}\,\int_\mathcal{M}  \left\langle A\wedge dA+\tfrac{2}{3} A\wedge A\wedge A\right\rangle
}{eq:ch9}
with specific boundary conditions imposed on the connection $A$ \cite{Henneaux:2010xg,Campoleoni:2010zq}. Choosing appropriate conditions at the boundary of the manifold $\mathcal{M}$ is often important in physics, but it is particularly crucial in Chern--Simons theories, since these theories exhibit no local physical degrees of freedom and all the physical states are boundary states. The same bulk theory \eqref{eq:ch9} can then lead to completely different boundary theories, depending on the chosen boundary conditions.

Our goal is now to find a suitable gauge algebra and suitable boundary conditions for the Chern--Simons connection $A$ so that we obtain flat space spin-3 gravity. Moreover, we need to construct the canonical boundary charges to verify the consistency of our boundary conditions. Finally, we intend to study the ensuing asymptotic symmetry algebra generated by the canonical boundary charges, in order to determine the symmetries of the putative field theory dual of flat space spin-3 gravity, as well as its central charges.

Let us start by proposing a spin-3 generalization of the $\textrm{iso}(2,1)$ algebra, with non-vanishing commutators
\begin{subequations}
\label{eq:FSHSG14}
\begin{align}
[ L_m, L_n]&=(m-n)L_{m+n}\\
[ L_m, M_n]&=(m-n)M_{m+n}\\
[ L_m, U_n]&=(2m-n)U_{m+n}\\
[ L_m, V_n]&=(2m-n)V_{m+n}\\
[ M_m, U_n]&=(2m-n)V_{m+n} \label{eq:FSHSG5} \\
[ U_m, U_n]&=(m-n)(2m^2+2n^2-mn-8)L_{m+n}\\
[ U_m, V_n]&=(m-n)(2m^2+2n^2-mn-8)M_{m+n}
\end{align}
\end{subequations}
where $m,n=\pm2,\pm1,0$ in $U$ and $V$ generators and $m,n=\pm1,0$ in $L$ and $M$ generators. Note that $L$ and $M$ generate $\textrm{iso}(2,1)$, which fixes their commutation relations with each other. The requirement that $U$ and $V$ be spin-3 fields fixes their commutation relations with $L$. The last two commutators are motivated by the requirements that $U$ ($V$) be the spin-3 partner of $L$ ($M$) and of course also by consistency with the Jacobi-identities. The commutator \eqref{eq:FSHSG5} is crucial to make contact with the spin-3 Fronsdal equation \cite{Fronsdal:1978rb}. For details on the relation to Fronsdal's equation see \cite{Campoleoni:2010zq}, whose analysis generalizes straightforwardly to the flat-space limit.
In fact, the algebra above emerges in the same way from a contraction of 
the global part of two copies of the $W_3$ algebra 
(see e.g.~\cite{Bouwknegt:1992wg,deBoer:1995nu}) as the BMS algebra emerges from a contraction of two copies of the Virasoro algebra \cite{Barnich:2006av,Bagchi:2010zz,Bagchi:2012cy}. 

The appropriate non-degenerate bilinear form of the algebra \eqref{eq:FSHSG14} is
\eq{
\langle L_m,\, M_n\rangle=-\tfrac{1}{2}\eta_{mn}\qquad\langle U_m,\, V_n
\rangle=-\tfrac{3}{2}\mathcal{K}_{mn}
}{eq:angelinajolie}
where $\eta_{mn}$ given by
$\eta = \textrm{antidiag}\, (1, \,-\tfrac12, \,1)$ 
is proportional to the $sl(2)$ Killing form and the $sl(3)$ part is given by 
$\mathcal K = \textrm{antidiag}\, (12, \,-3, \,2, \,-3, \,12)$ 
both of which have non-zero entries only on the anti-diagonal.

The bulk action of our flat space spin-3 theory is the Chern--Simons action \eqref{eq:ch9} with the gauge field 
given by
\eq{
A = A^n_L \,L_n+A^n_M \,M_n+A^n_U \,U_n+A^n_V \,V_n\,.
}{eq:FSHSG1}
As mentioned before, the theory is essentially empty without imposing suitable boundary conditions.

We propose the following boundary condition on the gauge field \footnote{We arrived at these boundary conditions by an ``educated guess'' and showed their consistency afterwards, inspired by the analogue result for the spin-2 case \cite{Afshar:2013bla}; alternatively, one could utilize the methods by Barnich and Gonzalez to take the AdS results formulated in a suitable way and then extract the flat space limit \cite{Barnich:2013yka}.
}.
\begin{subequations}
\label{eq:FSHSG13}
\begin{align}
  A_L^{-1}&= -\tfrac{1}{4}\mathcal{M}\,d\phi	\qquad
  A_L^{1} = d\phi \\
  A_M^{-1}&=-\tfrac{1}{4}\mathcal{M}\,du+\tfrac{1}{2}\,dr -\big(\tfrac{1}{2}\mathcal{L} + \tfrac u4 \mathcal{M}^\prime\big) \,d\phi \\
  A_M^{0}&=r\,d\phi  \qquad \qquad
  A_M^{1}=du \\
  A_U^{-2}&= \tfrac{1}{2}\mathcal{V}\,d\phi \qquad\quad 
  A_V^{-2}=\tfrac{1}{2}\mathcal{V}\,du+\big(\mathcal{U} + \tfrac u2 \mathcal{V}^\prime\big)\,d\phi 
  \end{align}
\end{subequations}
All other connection components are required to vanish at the asymptotic boundary $r\to\infty$.
The state-dependent functions $\mathcal{L, \,M, \,U, \,V}$ are allowed to depend on the angular coordinate $\phi\sim\phi+2\pi$ only, and prime denotes derivatives with respect to $\phi$.
The boundary conditions \eqref{eq:FSHSG13} are then consistent with the variational principle $\de I=0$, gauge flatness $F=dA+A\wedge A=0$, and our desire to obtain asymptotically flat line-elements as solutions.
It is useful for later purposes to introduce the combinations
$\mathcal{N} = \mathcal{L} + \tfrac u2 \mathcal{M}^\prime$ and 
$\mathcal{Z} = \mathcal{U} + \tfrac u2 \mathcal{V}^\prime$.
The linear dependence on the retarded time $u$ in the functions $\mathcal{N, \,Z}$ is anticipated from the spin-2 precedent \cite{Barnich:2006av}.

The corresponding line-element takes the form 
\begin{align}
\label{eq:FSHSG18}
 ds^2&=-2 \eta_{mn} A^m_M A^n_M   -2\,\mathcal{K}_{mn} A^m_V A^n_V \nonumber \\
&=\mathcal{M}du^2-2dudr+2\mathcal{N}dud\phi + r^2d\phi^2 \,.
\end{align}
The generators $M_n$ correspond to translation generators in a light-cone basis, which is why they appear in the relation to the dreibein, together with their spin-3 partner $V_n$. From the AdS case it is known that higher spin generators can appear in the vielbein, which is then sometimes referred to as ``zuvielbein''.
Note that only the state-dependent functions $\mathcal M,\, \mathcal N$ appear in the line-element. The asymptotically flat background solution (in Eddington--Finkelstein gauge) is obtained for the choice $\mathcal M=-1$ and $\mathcal N=0$.
The line-element \eqref{eq:FSHSG18} followed from evaluating the quadratic Casimir of $sl(3)$. Similarly, the cubic Casimir yields the spin-3 field $\Phi$, 
\eq{
\Phi_{mnk} dx^m dx^n dx^k = 2{\mathcal V}\,du^3+4{\mathcal Z}\,du^2d\phi \,.
}{eq:lalapetz}
It depends explicitly on the spin-3 functions $\mathcal V$ and $\mathcal Z$.

Boundary condition preserving gauge transformations, $\delta_\epsilon A = d \epsilon + [A, \,\epsilon]$,  are determined by the algorithm explained in \cite{Afshar:2012nk} (for non-AdS higher spin gravity), which we follow here. (This approach was pioneered by Ba\~nados \cite{Banados:1994tn} for AdS spin-2 gravity, realizing the seminal Brown--Henneaux analysis \cite{Brown:1986nw} in the Chern--Simons formulation.)

After a lengthy calculation we find that they are generated by gauge parameters $\epsilon$ with the following decomposition in terms of Lie algebra generators
\begin{subequations}
\label{eq:FSHSG11}
\begin{align}
\epsilon_L^{1}&=\epsilon & \epsilon_M^{1}&=2\tau & \epsilon_U^{2}&=\chi  & \epsilon_V^{2}&=\kappa\\
\epsilon_L^{0}&=-\epsilon' & \epsilon_M^{0}&=r\epsilon-2\tau' & \epsilon_U^{1}&=-\chi' 	& \epsilon_V^{1}&=-\kappa'+2r\chi \nonumber
\end{align}
with arbitrary functions $\epsilon$, $\sigma$, $\chi$, $\rho$ of the angular coordinate $\phi$ and
\begin{align}
\epsilon_L^{-1}&=\tfrac{1}{2}\epsilon''-\tfrac{1}{4}\mathcal{M}\epsilon +12 \mathcal{V}\chi	 \\
\epsilon_M^{-1}&=-\tfrac{1}{2}r\epsilon'+\tau''-\tfrac{1}{2}\mathcal{M}\tau -\tfrac{1}{2}\epsilon\mathcal{N}+12\mathcal{V}\kappa +24\chi\mathcal{Z}  		
\\ 
\epsilon_U^{0}&= \tfrac{1}{2}\chi''-\tfrac{1}{2}\mathcal{M}\chi \\	
\epsilon_U^{-1}&=-\tfrac{1}{6}\chi'''+\tfrac{1}{6}\mathcal{M}'\chi +\tfrac{5}{12}\mathcal{M}\chi' \\
\epsilon_U^{-2}&=\tfrac{1}{24}\chi''''-\tfrac{1}{24}\mathcal{M}''\chi-\tfrac{7}{48}\mathcal{M}'\chi'-\tfrac{1}{6}\mathcal{M}\chi'' \nonumber \\
&\quad +\tfrac{1}{16}\mathcal{M}^2\chi +\tfrac{1}{2}\mathcal{V}\epsilon
\\ 
\epsilon_V^{0}&=\tfrac{1}{2}\kappa''-\tfrac{3}{2}r\chi'-\tfrac{1}{2}\mathcal{M}\kappa-\mathcal{N}\chi\\
\epsilon_V^{-1}&=-\tfrac{1}{6}\kappa'''+\tfrac{1}{2}r\chi''+\tfrac{1}{6}\mathcal{M}'\kappa +\tfrac{5}{12}\mathcal{M}\kappa'+\tfrac{1}{3}\mathcal{N}'\chi  \nonumber \\
&\quad +\tfrac{5}{6}\mathcal{N}\chi'-\tfrac{1}{2}r \mathcal{M}\chi
\\
\epsilon_V^{-2}&=\tfrac{1}{24}\kappa''''-\tfrac{1}{12}r\chi'''-\tfrac{1}{24}\mathcal{M}''\kappa -\tfrac{7}{48}\mathcal{M}'\kappa' -\tfrac{1}{6}\mathcal{M}\kappa'' \nonumber \\
&\quad -\tfrac{1}{12}\mathcal{N}''\chi-\tfrac{7}{24}\mathcal{N}'\chi' -\tfrac{1}{3}\mathcal{N}\chi''+\tfrac{1}{12}r\mathcal{M}'\chi \nonumber \\
& \quad +\tfrac{5}{24}r\mathcal{M}\chi'+\tfrac{1}{16}\mathcal{M}^2\kappa +\tfrac{1}{4}\mathcal{M}\mathcal{N}\chi+\mathcal{Z}\epsilon+\mathcal{V}\tau
\end{align}
where
$\tau=\sigma +\tfrac{u}{2}\epsilon'$ and $\kappa=\rho +u\chi'$.
\end{subequations}

The results above lead to the following variations of the state-dependent functions. 
\begin{subequations}
\begin{align}
\delta_\epsilon \mathcal{L}&= \epsilon\mathcal{L}'+2\epsilon'\mathcal{L}\\
\delta_\epsilon \mathcal{M}&= \epsilon\mathcal{M}'+2\epsilon'\mathcal{M}-2\epsilon'''\\
\delta_\epsilon \mathcal{U}&= \epsilon\mathcal{U}'+3\epsilon'\mathcal{U}\\
\delta_\epsilon \mathcal{V}&= \epsilon\mathcal{V}'+3\epsilon'\mathcal{V}\\
\delta_\sigma \mathcal{L}&= \sigma\mathcal{M}'+2\sigma'\mathcal{M}-2\sigma'''\\
\delta_\sigma \mathcal{U}&= \sigma\mathcal{V}'+3\sigma'\mathcal{V}\\
\delta_\chi \mathcal{L}&= -48\chi\mathcal{U}'-72\chi'\mathcal{U}\\
\delta_\chi \mathcal{M}&= -48\chi\mathcal{V}'-72\chi'\mathcal{V}\\
\delta_\chi \mathcal{U}&= -\tfrac{1}{12}\chi\mathcal{L}'''-\tfrac{3}{8}\chi'\mathcal{L}''-\tfrac{5}{8}\chi''\mathcal{L}'-\tfrac{5}{12}\chi'''\mathcal{L} \nonumber \\
&\quad +\tfrac{1}{3}\chi(\mathcal{M}\mathcal{L})'+\tfrac{2}{3}\chi'\mathcal{M}\mathcal{L}\\
\delta_\chi \mathcal{V}&= -\tfrac{1}{12}\chi\mathcal{M}'''-\tfrac{3}{8}\chi'\mathcal{M}''-\tfrac{5}{8}\chi''\mathcal{M}'-\tfrac{5}{12}\chi'''\mathcal{M}\nonumber\\
&\quad+\tfrac{1}{3}\chi\mathcal{M}\mathcal{M}'+\tfrac{1}{3}\chi'\mathcal{M}^2+\tfrac{1}{12}\chi^{(5)}\\
\delta_\rho \mathcal{L}&= -24\rho\mathcal{V}'-36\rho'\mathcal{V}\\
\delta_\rho \mathcal{U}&= -\tfrac{1}{24}\rho\mathcal{M}'''-\tfrac{3}{16}\rho'\mathcal{M}''-\tfrac{5}{16}\rho''\mathcal{M}'-\tfrac{5}{24}\rho'''\mathcal{M}\nonumber\\
&\quad +\tfrac{1}{6}\rho\mathcal{M}\mathcal{M}'+\tfrac{1}{6}\rho'\mathcal{M}^2+\tfrac{1}{24}\rho^{(5)}
\end{align}
\end{subequations}

Following again the algorithm described in \cite{Afshar:2012nk} we determine next the canonical boundary charges, whose Poisson brackets then generate the asymptotic symmetry algebra. In order to bring them into a nice form, we rescale various quantities: $\tilde{\chi} = -24 \chi$, $\tilde{\rho} = -12 \rho$, $\tilde{\mathcal{L}} = k/(4\pi) \mathcal{L}$, $\tilde{\mathcal{M}} = k/(4\pi) \mathcal{M}$, $\tilde{\mathcal{U}} = -(6k/\pi)\mathcal{U}$, $\tilde{\mathcal{V}} = -(6k/\pi)\mathcal{V}$ and drop the tilde afterwards to reduce notational clutter.
For the canonical charges $Q$ we obtain then the result
\eq{
Q=-\int d\phi\,\big[\epsilon\,\mathcal{L}+\sigma\,\mathcal{M}
-\tfrac32\, \chi\,\mathcal{U}-\tfrac32\,\rho\,\mathcal{V}\big] \,.
}{eq:FSHSG6}
The charges \eqref{eq:FSHSG6} are finite, integrable and conserved in (retarded) time, $\partial_u Q=0$, which means that our proposal \eqref{eq:FSHSG13} for flat space boundary conditions in spin-3 gravity turned out to be consistent. This is our first key result.

The asymptotic symmetry algebra is the Poisson bracket algebra generated by the boundary condition preserving gauge transformations that have non-trivial canonical charges. These Poisson brackets can be written in terms of Fourier modes $\mathcal{L}_m$, $\mathcal{M}_m$, $\mathcal{U}_m$, $\mathcal{V}_m$, defined by
$\mathcal{L}(\phi) = \tfrac{1}{2\pi}\, \sum_{m \in \mathbb{Z}} \mathcal{L}_m e^{-i m \phi}$,
and similarly for $\mathcal{M}$, $\mathcal{U}$ and $\mathcal{V}$, except that we include additional factors $i$ on the right hand sides in the definitions of the Fourier-components for $\mathcal{U}$ and $\mathcal{V}$.

The remaining steps follow again the algorithm of \cite{Afshar:2012nk}.
With the shift $\mathcal{M}_0\to \mathcal{M}_0 + \tfrac k2$ and suitable quantum modifications of central terms to satisfy the Jacobi identities in the presence of normal ordering $::$ we replace the Poisson brackets by commutators (notationally indicated by replacing calligraphic generators like $\mathcal{L}_n$ by non-calligraphic ones like $L_n$) and obtain finally the asymptotic symmetry algebra of flat space spin-3 gravity.
\begin{subequations}
 \label{eq:FSHSG10}
\begin{align}
[L_m, L_n] &= (m-n) L_{m+n} + \frac{c_L}{12} (m^3 - m) \delta_{m,-n}  \\
[L_m, M_n] &= (m-n) M_{m+n} + \frac{c_M}{12} (m^3 - m) \delta_{m,-n}  \\
[L_m, U_n] &= (2m-n) U_{m+n}  \\
[L_m, V_n] &= (2m-n) V_{m+n}  \\
[M_m, U_n] &= (2m-n) V_{m+n}  \\
[U_m, U_n] &= (m-n)(2 m^2 + 2 n^2 - m n -8) L_{m+n} \nonumber \\ & \!\!\!\!\!\!\!\!\!\!\!\!\!\!\!\!
+\frac{192}{c_M}  (m-n)\Lambda_{m+n}  - \frac{96\big(c_L + \tfrac{44}{5}\big)}{c_M^2}  (m-n)\Theta_{m+n} \nonumber \\
& \quad +\frac{c_L}{12} m(m^2-1)(m^2-4) \delta_{m,-n} \label{eq:FSHSG20} \\
[U_m, V_n] &= (m-n)(2 m^2 + 2 n^2 - m n -8) M_{m+n} \nonumber \\ &\quad
+ \frac{96}{c_M}  (m-n) \Theta_{m+n} \nonumber \\
& \quad +\frac{c_M}{12} m(m^2-1)(m^2-4) \delta_{m,-n} 
\end{align}
We used the definitions $\Theta_m=\sum_p M_p\, M_{m-p}$ and
$\Lambda_m=\sum_p\colon L_p\, M_{m-p}\colon -\tfrac{3}{10}(m+2)(m+3)M_m$.
\end{subequations}
Normal ordering is defined by
$\colon L_n\, M_m\colon = L_n\, M_m$ if $n<-1$ and
$\colon L_n\, M_m\colon = M_m\, L_n$ otherwise.

Our main result \eqref{eq:FSHSG10} is the flat space spin-3 generalization of the centrally extended BMS algebra in three dimensions \cite{Barnich:2006av}.
The dual field theory, if it exists, must then fall into representations of the algebra \eqref{eq:FSHSG10}. To the best of our knowledge this algebra has not appeared in the literature before. 

We have also calculated the values of the two central charges appearing in the algebra \eqref{eq:FSHSG10} and find \footnote{We have allowed for a non-zero central charge $c_L$ in \eqref{eq:FSHSG10} to facilitate generalizations to other spin-3 theories. Note also that, as expected, the center-less subalgebra coincides precisely with the algebra \eqref{eq:FSHSG14}.}
\eq{
c_L = 0\qquad c_M = 12k\,.
}{eq:FSHSG15}
These are the same results as for the corresponding central charges in spin-2 gravity \cite{Barnich:2006av}. 

We conclude with several physical and mathematical remarks that point to future directions and provide additional information on the interpretation of our results.

It is straightforward to show that the algebra \eqref{eq:FSHSG10} arises as an \.In\"on\"u--Wigner contraction (``ultra-relativistic boost'') of two copies of the infinite-dimensional $W_3$ algebra (with Virasoro generators ${\cal L}$, spin-3 generators ${\cal W}$, central charge $c$ in one copy and $\cal\bar L$, $\cal\bar W$, $\bar c$ in the other copy), including the quantum shift of $44/5$ in the commutator \eqref{eq:FSHSG20}. The relations between the respective generators are
$L_n = {\cal L}_n - {\cal\bar L}_{-n}$, $ M_n = \tfrac1\ell\big({\cal L}_n + {\cal \bar L}_{-n}\big)$, $U_n = {\cal W}_n - {\cal\bar W}_{-n}$ and $V_n = \tfrac1\ell\big({\cal W}_n + {\cal\bar W}_{-n}\big)$.  
The central charges are related by $c_L=c-\bar c$ and $c_M=\tfrac1\ell\,(c+\bar c)$, and the limit $\ell\to\infty$ is taken when contracting to the flat-space algebra \eqref{eq:FSHSG10}. This is completely analog to what happens in the spin-2 case in the contraction from two copies of the Virasoro algebra to the Galilean conformal algebra \cite{Bagchi:2009my,Bagchi:2010zz,Bagchi:2012cy}.

The spin-2 generators $M_n$ and $L_n$ generate so-called super-translations and super-rotations, respectively \cite{Ashtekar:1996cd,Barnich:2006av}. The generators $V_n$ ($U_n$) are then spin-3 analogues of super-translations (super-rotations). The usual vacuum in the dual field theory is annihilated by all generators with non-negative index or belonging to the center-less subalgebra. Perturbative physical states in the dual field theory are then generated by descendants of the vacuum with (sufficiently) negative index. For instance, $L_{-n}|0\rangle$ with $n>1$ generates usual Virasoro descendants of the vacuum, and similarly for $U_{-n}|0\rangle$, $M_{-n}|0\rangle$, $V_{-n}|0\rangle$ and arbitrary combinations thereof.
It is not clear if the algebra \eqref{eq:FSHSG10} has unitary representations for $c_M\neq 0$. This is an important open issue which we defer to future studies.

It is interesting to check whether there are other flat space spin-3 theories that lead to different results for the central charges, in particular a  spin-3 analogue to flat space chiral gravity \cite{Bagchi:2012yk} where $c_L\neq 0$, $c_M=0$. The limit $c_M\to 0$ requires an additional scaling $U_n\to c_M U_n$ in the algebra \eqref{eq:FSHSG10}. After taking the limit the contracted algebra has the same commutators as \eqref{eq:FSHSG10}, with the following exceptions \footnote{This contraction resembles the one for the $W_3$ algebra considered by Hull and Palacios \cite{Hull:1992gx}.}: $[L_m,\, M_n] = (m-n)\,M_{m+n}$, $[M_m,\, U_n]= 0$  and
 $[U_m,\,U_n] \propto [U_m,\, V_n] = 96 (m-n)\Theta_{m+n}$.
The spin-2 super-translation generators and all spin-3 generators consistently can be set to zero, $M_{-n}|0\rangle = U_{-n}|0\rangle = V_{-n}|0\rangle = 0$. Consequently, for $c_M=0$ the only (perturbative) physical states are the Virasoro descendants of the vacuum, $L_{-n}|0\rangle$ with $n\geq 2$.
Thus, the limit $c_M\to 0$ eliminates the higher spin degrees of freedom and gets us from a flat-space higher spin theory to a spin-2 theory.

Let us now briefly address some non-perturbative solutions to our theory. Even with spin-3 charges turned on, the Ba\~nados--Teitelboim--Zanelli black holes \cite{Banados:1992wn} continue to be solutions of the higher spin AdS$_3$ theory. Similarly, flat space cosmologies \cite{Cornalba:2002fi,Cornalba:2003kd,Barnich:2012xq,Bagchi:2012xr} 
continue to be solutions to spin-3 gravity in 3-dimensional flat space. If we choose the state-dependent functions in the line-element \eqref{eq:FSHSG18} as ${\cal L}= -\hat r_+ r_0$, ${\cal M}=\hat r_+^2$ and make elementary coordinate transformations then we reproduce precisely the flat space cosmology line-element in the notation of \cite{Bagchi:2013lma}.

It is also possible to construct flat space solutions similar to the Gutperle--Kraus higher spin AdS$_3$ black hole \cite{Gutperle:2011kf} by switching on marginal or relevant deformations in the spin-3 field. This would amount to turning on a chemical potential for the higher spin charges. These spin-3 cosmological solutions solve the bulk equations of motion but violate our present boundary conditions. We present such an example below. We turn on $A_U^a = V^a d\phi$, $A_V^a = V^a du + [(a+3)V^{a+1}r/2 + (V^a)^\prime u + U^a]d\phi$ where $V^a$ and $U^a$ are arbitrary functions of $\phi$ and the integer index $a$ runs from $-2$ to $2$. The undeformed solutions consistent with the boundary conditions \eqref{eq:FSHSG13} have $V^{-2}={\mathcal V}/2$, $U^{-2}=\mathcal U$ and $V^a=U^a=0$ otherwise. The quantities $V^2$ and $U^2$ are sources for spin-3 charges.

It would be of interest to study phase transitions between flat space spin-3 gravity and these flat space cosmologies, analogous to the ones arising in the spin-2 case \cite{Bagchi:2013lma}. It might turn out that like in the AdS example where the Hawking--Page transition is smoothened out \cite{Banerjee:2012aj}, higher spins also wash away the flat space spin-2 transitions.
Moreover, there could be a spin-3 generalization of the Cardy-like formula derived in Refs.~\cite{Barnich:2012xq,Bagchi:2012xr}.
This should be checked by combining the flat space spin-2 analysis with the AdS spin-3 analysis \cite{Gutperle:2011kf}.
Perhaps new light can be shed on the ongoing spin-3 entropy discussion (see e.g.~\cite{Gutperle:2011kf,Kraus:2011ds,Perez:2012cf,Campoleoni:2012hp,Perez:2013xi,deBoer:2013gz,Kraus:2013esi,deBoer:2013vca,Ammon:2013hba}) by applying the various approaches to spin-3 flat space cosmologies (with spin-3 
chemical potentials).

It should be straightforward to generalize the results of this paper to arbitrary 3-dimensional higher spin theories, including the limit of infinite spins, the so-called $\textrm{hs}_\lambda$ theory or perhaps to the massive Prokushkin--Vasiliev theory \cite{Prokushkin:1998bq}.

Finally, it would be interesting to understand the relation of higher spin theory to string theory and its tensionless limit especially in the present 3-dimensional flat space case as the symmetries of the world-sheet turn out to be the spin-2 sub-sector of \eqref{eq:FSHSG10} \cite{Bagchi:2013bga}.

\acknowledgments

We are grateful to X.~Bekaert, J.~de~Boer, A.~Campoleoni, A.~Castro, S.~Detournay, M.~Gaberdiel, M.~Gary, M.~Henneaux, J.~Jottar, E.~Kiritsis, I.~Klebanov, A.~Perez, E.~Perlmutter, R.~Rashkov, S.-J.~Rey, M.~Riegler, J.~Simon, W.~Song, P.~Sundell, D.~Tempo, R.~Troncoso and M.~Vasiliev for discussions.
HA, DG and JR were supported by the START project Y~435-N16 of the Austrian Science Fund (FWF) and the FWF projects I~952-N16 and I~1030-N27. AB is supported by the Engineering and Physical Sciences Research Council, UK and an INSPIRE award by the Department of Science and Technology, India. RF would like to acknowledge the hospitality of the Institute for Theoretical Physics, TU Vienna during the first stage of this work.
AB and DG thank the Center for Theoretical Physics (CTP) at MIT for hospitality during the early stages of this work.
HA, AB and DG thank the Galileo Galilei Institute for Theoretical Physics for the hospitality and the INFN for partial support during this work.
DG thanks the Asia Pacific Center for Theoretical Physics (APCTP) and the Centro de Ciencias de Benasque for hospitality during the completion of this work.


\paragraph{Note added in proofs:}
A few days after submission of our Letter to the arXiv an e-print appeared by Gonzalez et al.~\cite{Gonzalez:2013oaa} that also considers asymptotically flat spacetimes in three-dimensional higher spin gravity. Where applicable, our results agree with each other.

\enlargethispage{0.4cm}

\end{document}